\documentclass[11pt]{article}
\usepackage{times}
\usepackage{graphics}
\usepackage{amsmath}
\usepackage{amssymb}
\usepackage{ifthen}
\usepackage{calc}
\usepackage{hyperref}
\usepackage{color}

\newcommand{\vio}[1]{\ensuremath{\widehat{#1}}}
\newcommand{\flow}[1]{\ensuremath{\widehat{#1}_f}}
\newcommand{\up}{\rule{0.0in}{0.18in}}

\newcommand{\nlog}[2]{\ensuremath{\Theta\left(n\ifthenelse{\equal{#1}{1}}{}{^{#1}}\ifthenelse{\equal{#2}{0}}{}{\log\ifthenelse{\equal{#2}{1}}{}{^{#2}} n}\right)}}

\newtheorem{theorem}{Theorem}
\newtheorem{corollary}[theorem]{Corollary}

\definecolor{linkcolor}{rgb}{0.1,0.1,0.8}    

\begin{document}

\begin{center}
{\Large \textbf{$L_p$ Isotonic Regression Algorithms Using an $L_0$ Approach}}
\bigskip

{\large Quentin F. Stout}\\[\medskipamount]
qstout@umich.edu\\
www.eecs.umich.edu/{\,\raisebox{-0.5ex}{\textasciitilde}qstout/}

\end{center}
 \bigskip

\noindent \textbf{Abstract:~}
Significant advances in flow algorithms have changed the relative performance of various approaches to algorithms for $L_p$ isotonic regression.
We show a simple plug-in method to systematically incorporate such advances, and advances in determining violator dags, with no assumptions about the algorithms' structures.
The method is based on the standard algorithm for $L_0$ (Hamming distance) isotonic regression (by finding anti-chains in a violator dag), coupled with partitioning based on binary $L_1$ isotonic regression.
For several important classes of graphs the algorithms are already faster (in O-notation) than previously published ones, close to or at the lower bound, and significantly faster than those implemented in statistical packages.
We consider exact and approximate results for $L_p$ regressions, $p=0$ and $1 \leq p < \infty$, and a variety of orderings.
\medskip 

\bigskip

\section{Introduction}

There are many scenarios when one expects that an attribute of objects increases as a function of other attributes.
For example, the expected maximum grade level achieved by a child is likely to be an increasing function of their mother's maximum grade level and father's maximum grade level.
No assumptions are made about the relationship between a lower mother's grade level and higher father's grade level versus a higher mother's grade level and lower father's.
More generally, let $V$ be a set with a partial order $\prec$, and let $f$ be a function on $V$. 
$f$ is \textit{isotonic} iff for all $u,v \in V$, if $u \prec v$ then $f(u) \leq f(v)$.
Thus we expect that the child's maximum grade level is an isotonic function of its mother's maximum grade level and its father's, 
but often the data collected isn't quite isotonic, so regression is needed to determine the underlying relationship.

Given arbitrary real-valued functions $f,g$, and nonnegative weight function $w$ on $V$ and a $p$, where $p=0$ or $1 \leq p \leq \infty$, their weighted $L_p$ distance, denoted $||f-g||_p$, is 
$$
\begin{array}{ll}
  \sum_{v \in V} w(v) \cdot \mathbf{1} (f(v) \neq g(v))   & p = 0 \\
  \left(\sum_{v \in V} w(v) \cdot |f(v) -g(v)|^p\right)^{1/p} & 1 \leq p < \infty \smallskip \\
  \max_{v \in V} w(v) \cdot |f(v) - g(v)|                     & p = \infty
\end{array}
$$
$L_0$ is not a true norm and is is also known as the \textit{Hamming distance}, 0-1 loss, or Kronecker delta loss.
\textit{Unweighted function} means $w(v) =1$ for all $v \in V$.

$f^\prime$ is an $L_p$ \textit{isotonic regression} of $f$ iff it minimizes $||f-g||_p$ among all isotonic functions $g$.
Isotonic regressions are unique for $1 < p < \infty$, but not necessarily for $p=0, 1, \infty$.
Many different terms are applied to various versions of this.
For example, in mathematical analyses often monotonic is used instead of isotonic.
$L_0$ isotonic regression is sometimes called monotonic relabeling, and the distance is also known as Hamming distance, 0-1 distance or 0-1 loss.
$L_1$ regression is sometimes called median regression, the least absolute deviations, or sum of deviations error.
The $L_0$ regression error of the optimal isotonic regression is sometimes called the ``distance to monotonicity''.
$L_2$ is the (sometimes unstated) most common norm for isotonic regression.
Isotonic regression is a non-parametric shape-constrained regression.

The paper concentrates on algorithms, not on any specific application.
Some relevant $L_0$ applications appear in \cite{Brabantetal,LearnRelab_Cano18,DuivFeeldersNNclass,Relab_Feelders06,OrdinalMonotone,LinearSingle,PTFuzzy,Truth,Relab_PijlsPot14,Relab_Radeeta09,Relab_Radetal12,RULEM17}, and for $L_1$ and $L_2$ the applications are far too numerous to list.
We also omit aspects such as the rate of convergence of isotonic estimators.

Often the partial order is given via a dag which is connected (for disconnected ones the algorithms would be applied on each component separately), having $n$ vertices and $m$ directed edges.
Efficient algorithms for determining isotonic regressions depend on the metric and the underlying dag.
$L_\infty$ algorithms are quite different than those for $L_p$, $ p < \infty$, and won't be discussed.
For $L_0$, $L_1$, and $L_2$ the fastest previous algorithms utilize variations on maximum flows and linear programming.
Those for $L_0$  use minimum flow algorithms and take $\Theta(n^3)$ time~\cite{Relab_Feelders06,Relab_PijlsPot14,Relab_Radeeta09};
for $L_1$ a linear programming approach is used, taking $\Theta(nm+n^2 \log n)$ time~\cite{AHKW}; and for
$L_2$ a maximum flow approach is used, taking $\Theta(nm \log(\frac{n^2}{m}))$ time~\cite{HochQuey}.
There do not seem to be any published $L_p$ algorithms for other values of $p < \infty$, other than approximation algorithms~\cite{AhujaOrlin,PolyLp,Yale,QMultidim,Strom}.

Recent advances in flow algorithms change some of the relative timing of these approaches, though under the assumption that the function values and weights are integers.
If $U$ is the maximum integer value and weight then 
the BLLSSSW minimum cost flow algorithm~\cite{Flow_BLLSSSW} takes $\tilde{O}\big(m+n^\frac{3}{2} \log U\big)$  time
and the CKLPPS algorithm~\cite{Flow_CKLPPS} takes $O\big(m^{1+o(n)} \log U\big)$ time, both with high probability.
The algorithms mentioned above are deterministic and give exact results, but here we consider the BLLSSSW and CLLPPS algorithms since they achieve a significantly faster expected time with high probability, and we can set $U$ to achieve whatever precision is necessary.
There is intense work in this area, and the wikipedia page~\cite{wiki} is quickly updated when faster flow algorithms are found.

Our results are stated in terms of general flow algorithms, and then specific times are given when one of the above flow algorithms is used, though other flow algorithms could be used as well as new advances occur, or one may use the analyses to understand the effect of using slower, but simpler, flow algorithms.
We use $T(\mathcal{F}(n,m,U))$ to denote the time of the flow algorithm of interest for a graph of $n$ vertices, $m$ edges, and maximum edge capacity $U$.
For a dag $G$ we use $T(\mathcal{V}(G))$ to indicate the time to find a violator dag of $G$ (defined in Section~\ref{sec:L0}).
Here too this is stated in general terms, and then the currently fastest known algorithms are referred to.
One important aspect of the work here is that one does not need to know the internal structure of the algorithms and does not need to modify them.
Instead we merely use them as black boxes, which should make it much easier for a wider range of researchers to apply and for updates to be available.

In Section~\ref{sec:background} we explain the approach that will used.
In Section~\ref{sec:L0} the time of the approach for $L_0$ isotonic regression is analyzed and the role of $\mathcal{F}$ and $\mathcal{V}$ will be made clear.
Section~\ref{sec:partition} introduces \textit{partitioning}, used for $L_1$ regression, and shows its role in approximate and exact $L_p$ regression.
Section~\ref{sec:multidim} shows important classes of dags, such as points in $d$-dimensional space, where the violator graph can be found more quickly than the general case.
Section~\ref{sec:final} contains final remarks and tables comparing the results here to the fastest previously known algorithms.

\section{Background} \label{sec:background}

Often isotonic regression algorithms are described in terms of violating pairs and violator dags.
Given a dag $G=(V,E)$ and a real-valued function $f$ on $V$, vertices $u, v \in V$ are a \textit{violating pair} if $u \prec v$ and $f(u) > f(v)$,
i.e., they violate the isotonic requirement.
Using $u \prec_v v$ to denote that $u$ and $v$ are a violating pair, it is easy see that $\prec_v$ defines a partial order on $V$.
A dag $\vio{G}=(\vio{V},\vio{E})$, is a \textit{violator dag} of $G$ iff $V \subset \vio{V}$ and for any $u,v \in V$ there is a path from $u$ to $v$ in $\vio{G}$ iff $u \prec_v v$.
There may be many violator dags for a given $G$ and $f$.
For example, if $\vio{G}$ has $V=\vio{V}$ and there is an edge $(u,v) \in \vio{E}$ iff $u$ and $v$ are a violating pair in $G$, then $\vio{G}$ is the transitive closure of the violating pairs.
The edges of the smallest subgraph of the transitive closure which is still a violator dag form the transitive reduction.
In Section~\ref{sec:multidim} we show that sometimes it is useful to have $\vio{V}$ contain additional vertices not in $V$.

We leave the algorithm to create the violator dag  unspecified, and then show the time for specific algorithms.
This generality will only be used in Section~\ref{sec:multidim}, though might be useful in other circumstances as well.
We usually assume that the standard approach is used, creating the transitive closure of $G$ and then removing edges that are not a violating pair.
For arbitrary dags of $n$ vertices the best possible time for finding the transitive closure or the transitive reduction is linear in the time for Boolean matrix multiplication~\cite{Matrix_Reduction,Matrix_Fischer,Matrix_Furman,Matrix_Munro}, while for specific dags, such as linear orders, the time can be $O(nm)$.
We will use $T(\vio{G}) = O(\min\{nm, n^\omega\})$ for general orderings,
where $\omega$ is such that Boolean matrix multiplication can be computed in $O(n^\omega)$ time. 
There are values of $\omega < 2.4$, but current algorithms achieving this are galactic.
One could use a practical algorithm, such as Strassen's, for which the time is $\Theta(n^{\lg 7 \approx 2.81})$.

The overall structure of the approach used here is:
\begin{itemize}

\item Using topological sort, determine all vertices which are not in any violating pair and remove them and all adjacent edges. The time for this is $\Theta(m)$ and will be ignored from now on, and $G$, $n$ and $m$ will refer to the remaining graph. For all vertices removed at this stage their value in the isotonic regression will be the original function value.
 
\item Use a violator dag and a minimum cost flow algorithm to find an $L_0$ isotonic regression~Sec.~\ref{sec:L0}
    \begin{itemize}
    \item For some important orderings there are fast algorithms for finding small violator graphs, use these instead to improve the overall time~\ref{sec:multidim}.
    \end{itemize}
\item View $L_0$ monotonic regression when there are only two labels as $L_1$ monotonic regression when the only function values are $\{0,1\}$.
Based on partitioning~\cite{QPartition}, use $\{0,1\}$-valued $L_1$ isotonic regressions to find isotonic regressions on data which has integer values and weights~\ref{sec:L1}.
Incorporating scaling, the regression is
   \begin{itemize}
       \item For $L_1$ and $L_2$: exact .
      \item For $L_p$, $1 < p < \infty$: an approximation with specified error bound.
   \end{itemize}
\end{itemize}
\noindent
There are algorithms for some simple orderings, such as a linear ordering, which do not use this approach~\ref{sec:multidim}.
The tables in Table~\ref{tab:oldvsnew} include these orderings for completeness, and for $L_2$ show that the algorithms most often implemented are not optimal.

Because partitioning is being used, and some algorithmic approaches exploit the fact that the $L_1$ regression is $\{0,1\}$-valued,
we use $T(\mathcal{F}_2(n,m,U))$ to denote the time when the function is $\{0,1\}$-valued.

\section{$L_0$ Isotonic Regression} \label{sec:L0}

Most of this section is a rephrasing and condensation of the Background section of~\cite{QL0}, which is based on the work of~\cite{Relab_Feelders06,Mohring,Relab_PijlsDilworth13,Relab_Radetal12}, for $L_0$ isotonic regression.
The fundamental algorithm for finding an $L_0$ isotonic regression is given in Figure~\ref{fig:general}, 
a slightly modified version of a figure in~\cite{QL0}.
Note that in $L_0$ there is no need for a metric on the function values, merely that they are linearly ordered, while for $L_p$, 
$1 \leq p \leq \infty$, a metric is required.

\begin{figure} 
\begin{enumerate}

\item Create a violator dag $\vio{G}=(\vio{V},\vio{E})$ of $G$.
\item Find an antichain $C$ of $\vio{G}$ of maximum weight, where the weight of $C$ is $\sum_{v \in C} w(v)$:
  \begin{enumerate}
     \item Create a flow graph $\flow{G}$ from $\vio{G}$.
     \item Find a minimum flow on $\flow{G}$ and use this to determine $C$.
   \end{enumerate}
\item Determine an isotonic regression $f^\prime$ on $G$:
$$ 
f^\prime(v) = 
          \left\{ 
                    \begin{array}{ll}
                     f(v)   & v \in C \\
                     \max\{f(w): w \prec v, w \in C\}   & \mathrm{if~there~is~a~predecessor~of~} v \mathrm{~in~} C \\
                     \min\{f(w): v \prec w, w \in C\}   & \mathrm{otherwise}
                     \end{array}
            \right.
$$
\end{enumerate}

\caption{Optimal $L_0$ monotonic relabeling $f^\prime$ of label function $f$ on $G=(V,E)$ with order $\prec$  (see \cite{Relab_Feelders06,Mohring,Relab_PijlsDilworth13,Relab_Radetal12})  }
\label{fig:general}
\hrulefill
\end{figure}

The role of $C$ in Figure~\ref{fig:general} is that any vertices which are an anti-chain (i.e., there are no directed paths connecting any pair) in $\vio{G}$ can be used as vertices where $f = f^\prime $ in the isotonic regression.
In fact, given any set of vertices $D$ where $f$ is unchanged they can only be part of an isotonic function if $D$ is an anti-chain.
Thus picking an anti-chain of maximum weight yields an optimal $L_0$ isotonic regression, if the remaining values are chosen to create an isotonic function.
That is what step 3 does: if $v$ has predecessors in $C$ then $f^\prime(v)$ must be no smaller than $f^\prime(w)=f(w)$ for any predecessor of $v$ in C, and similar it must be no larger than $f(w)$ for any successors in $C$.
Note that if $v \not\in C$ then $v$ must have successors or predecessors, but not both, in $C$, since otherwise it could be added to $C$ and increase the anti-chain.
There are other possibilities for values of $f^\prime$ on $V \setminus C$ but they aren't needed here.

\begin{figure}

\centerline{\scalebox{0.85}{\includegraphics{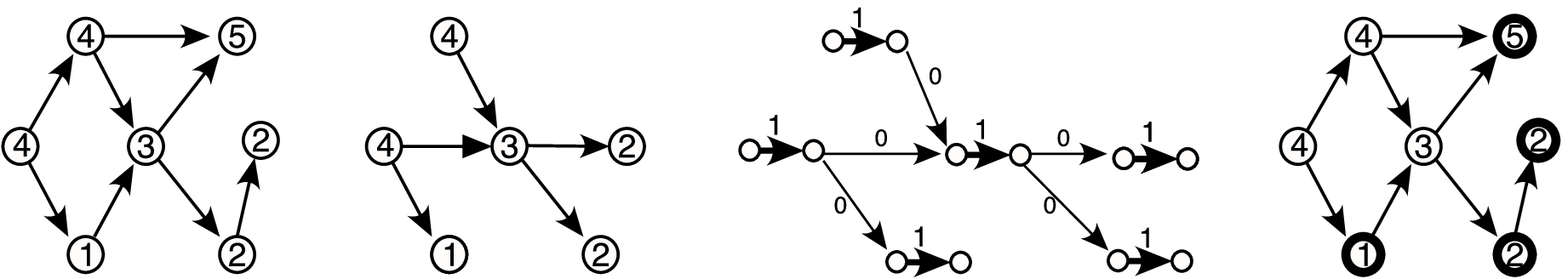}}}

\caption{Label function on unweighted dag $G$,~ one of its violator dags $\vio{G}$,~ flow graph $\flow{G}$,~ resulting $C$}
\label{fig:flowgraph}

\hrulefill
\end{figure}

A flow graph is used to find $C$.
This is illustrated in Figure~\ref{fig:flowgraph}, which is copied from~\cite{QL0}.
The flow graph is straightforward to construct from the violator dag $\vio{G}$, replacing each vertex $v$ with a pair $v_\mathrm{in}$ and $v_\mathrm{out}$, with an edge from $v_\mathrm{in}$ to $v_\mathrm{out}$.
There is an edge from $w_\mathrm{out}$ to $v_\mathrm{in}$ iff $(w,v)$ is an edge in $\vio{G}$.
Edges of the form $(v_\mathrm{in},v_\mathrm{out})$ have required flow at least 1 (or $w(v)$ in the weighted case) and all other edges have weight 0.
A minimum cost flow is found where the flow on each edge is at least the edge's requirement, and this is used to find a maximum cut $C$.
This is essentially the same as max-flow min-cut, but first one needs to find a feasible flow on $\vio{G}$ (one meeting each edge's requirement) and then minimize it.
This is done by subtracting the requirement from the flow, using these values as the upper capacities on the edges of $\vio{G}$ and finding a maximum flow on this, and then subtracting this maximal flow from the feasible flow to get the final flow.

Further explanation is in~\cite{Relab_Feelders06,Mohring,Relab_Radetal12}, but here the important part is that the time is dominated by the time to create $\vio{G}$, and the time of the minimum cost flow algorithm.
Finding a feasible flow can easily be done using a forward and backward topological sort, creating an excess of at most $nU$, where $U$ is the maximum weight on any vertex.
Thus the dominating factor for the flow is the max flow adjustment, $\mathcal{F}(\vio{n},\vio{m}, nU)$.

\newcounter{beana}

Note that for a violator graph $\mathcal{V}(G)$, $\vio{m}$ may be $\binom{n}{2}$ even when the number of edges in $G$ is $n-1$
(consider a linear order where the values are in decreasing order).
Thus when the transitive closure approach is used to create the violator the worst-case time over all dags with $n$ vertices is 
$O(n^{\omega} + \mathcal{F}(n,n^2,nU))$.
If the violator dag is created by determining the transitive closure, and the flow algorithm used is the faster of the BLLSSSW or CLLPPS flow algorithms then:

\newcounter{thmlistcounter}

\begin{theorem} \label{thm:L0L1}
Given a dag $G=(V,E)$ with weighted function $(f,w)$ on $V$ where the weights are integers in $[0,U]$:
\begin{list}
{\alph{thmlistcounter})}{\usecounter{thmlistcounter}\setlength{\leftmargin}{0.3in}}

\item an $L_0$ isotonic regression can be found in $O(\mathcal{V}(G) + \mathcal{F}(\vio{n},\vio{m}, nU))$ time. 
For arbitrary dags currently the fastest is
$$\tilde{O}\left(\min\{nm, n^\omega\} + \vio{m} \log U\right)$$
using the BLLSSSW algorithm~\cite{Flow_BLLSSSW}, and if a violator graph is given is
$$\tilde{O}\left(\min\{\vio{m}^{1+o(n)},\, \vio{m}+\vio{n}^{1.5}\}\log U\right)$$
using the faster of the BLLSSSW~\cite{Flow_BLLSSSW} and CKLPPS~\cite{Flow_CKLPPS} algorithms.
All times are with high probability.

\item if the function values are integers with $\ell$ different values then an $L_1$ isotonic regression can be found in
 $O\left(\mathcal{V}(G) + \mathcal{F}_2(\vio{n},\vio{m}, U) \!\cdot\! \log \ell ) \right)$ time.
 For arbitrary dags currently the fastest  is
 $$\tilde{O}\left(\min\{nm, n^\omega\} + \vio{m} \log \ell \log U\right)$$
 using the BLLSSSW algorithm~\cite{Flow_BLLSSSW}, and if a violator graph is given is
 $$\tilde{O}\left(\min\{\vio{m}^{1+o(n)},\, \vio{m}+\vio{n}^{1.5}\}\log \ell \log U\right)$$
using the faster of the BLLSSSW~\cite{Flow_BLLSSSW} and CKLPPS~\cite{Flow_CKLPPS} algorithms.
All times are with high probability.
\end{list}

\end{theorem}
\medskip
\noindent 
Given the transitive closure one can always create a violator graph with the same vertices as the original and the directed edges are always between violating pairs.
In this violator graph $\hat{n} =n$ and $\hat{m} \leq \binom{n}{2}$.

For both $L_0$ and $L_1$ the worst-case time over all dags of $n$ vertices and $m$ edges is $O\left(\min\{nm,n^\omega\}\right)$, i.e., finding a violator graph is the bottleneck.
Section~\ref{sec:multidim} gives classes of graphs for which this is no longer true, and
the resulting times are summarized in Table~\ref{tab:oldvsnew}.
Claim \textit{i}) was proven above, and \textit{ii}) will be shown in Section~\ref{sec:L1}.
Previous results only showed that an $L_0$ isotonic regression can be found in $\Theta(n^3)$ time even when a violator dag was given~\cite{Relab_Feelders06,Relab_PijlsPot14,Relab_Radeeta09}.
An $L_1$ isotonic regression algorithm taking $\Theta(nm + n^2 \log n)$ time appeared in~\cite{AHKW}, which also takes $\Theta(n^3)$ time in the worst case over all dags of $n$ vertices.

\section{Partitioning}  \label{sec:partition}

When there are only 2 labels $L_0$ isotonic regression is the same as $L_1$ isotonic regression when the values are in $\{0,1\}$, i.e., maximizing the weight of the unchanged labels is the same as minimizing the weight of those  that do change.
Thus an $L_0$ isotonic regression can be found by using an $L_1$ isotonic regression algorithm.
Here we reverse that. 
We start with the $L_0$ approach of building a violator dag then finding a flow and use that to construct $L_1$ regression where the function values are in $\{0,1\}$.
Then we use \textit{partitioning} to find $L_1$ regressions for weighted functions with arbitrary values, as well as regressions for $L_p$, $1 < p < \infty$.
It's important to note that while $L_0$ for 2 labels is essentially the same as $L_1$ for $\{0,1\}$, it is not in general true for $k$ labels and $\{0, \ldots, k-1\}$.
As far as the author knows, in general one cannot recursively apply partitioning to find $L_0$ isotonic regressions.

\subsection{$\mathbf{L_1}$ Isotonic Regression} \label{sec:L1}
  
In this section we prove claim \textit{ii}) of Theorem~\ref{thm:L0L1}.

To expand to an $L_1$ algorithm for arbitrary values, not just binary ones, we use an observation in~\cite{QPartition} (and essentially in~\cite{HochQuey}).
Suppose no values of $f$ are in $(a,b)$, $a<b$.
Create a new weighted function $g$ on $V$ where $g(v) = a$ if $f(v)\leq a$, and $g(v) = b$ if $f(v) \geq b$.
The weight at $v$ is the same as its original weight, $w(v)$.
There is always an $L_1$ optimal isotonic regression where the regression values are the initial function values, so there is an optimal regression $g^\prime$ of $g$ where every value of $g^\prime$ is $a$ or $b$.
Further,~\cite{QPartition} showed that for any such $g^\prime$, there is an $L_1$ isotonic regression $f^\prime$ of $f$, where $f^\prime(v) \leq a$ if $g^\prime(v) = a$, and $f^\prime(v) \geq b$ if $g^\prime(v) =b$.
Using the original weights as the weights in the binary regression represents the increase in error that would occur if $g(v) = a$ but $g^\prime(v)=b$.
In this case $[g^\prime(v)\!-\!f(v)] - [g(v)\!-\!f(v)]$ is $b-a$, so the error increased by $(b-a)w(v)$.
The same is true if $g(v)=b$ but $g^\prime(v)=a$.
Since the change in errors is always proportional to $b-a$ we can divide by that term and minimize the increase in error by using the original weights.
This is just weighted $\{0,1\}$-valued $L_1$ regression shifted, so here it is just referred to as $\{0,1\}$-valued $L_1$ regression.

Let $V_a$ be the vertices where $g^\prime = a$ and $V_b$ the vertices where it is $b$, and let $G_a = (V_a,E_a)$ and $G_b = $ $(V_b, E_b)$,
where $E_a$ contains those edges with both endpoints in $V_a$, and similarly for $E_b$.
Edges with one endpoint in $a$ and the other in $b$ are discarded.
No vertex of $V_b$ precedes any vertex of $V_a$, and the final regression values of each set only depend on the weighted function values in that set, so the regressions on $G_a$ and $G_b$ can be computed independently.

The way this is applied is that the initial $a$ and $b$ values are the medians of the original function values (if there are 2 medians), or $a$ is the median and $b$ the smallest value larger than $a$.
Then $G_a$ is partitioned using the first quartile of the original function values (in increasing order) and $G_b$ the third quartile.
This halving process continues until, in every subgraph, the splitting values are the only two values remaining in their subgraph.
There are at most $\Theta(\log \ell)$ iterations if there are only $\ell$ different function values.

This finishes the proof of Theorem~\ref{thm:L0L1}~ \textit{ii}), and hence finishes the proof of the theorem.
$\Box$

\subsection{Approximate and Exact $\mathbf{L_p}$ Regression, $\mathbf{1 < p < \infty}$} \label{sec:approx}

Various approximation algorithms for $L_p$ isotonic regression have appeared, with~\cite{AhujaOrlin,PolyLp,Strom} considering linear dags and~\cite{Yale} considering arbitrary ones.
Ones based on partitioning using the $L_1$ algorithm appear in~\cite{QPartition}.
There the basic idea is that partitioning is correct for general $L_p$, $1 \leq p < \infty$, so binary $L_1$ regression can be used in a similar fashion as for general $L_1$ regression.
However, it is no longer true that the regression values can always be chosen to be values of the original function.
E.g., for a linear order of 2 vertices, with unweighted values 1, 0, the unique $L_p$ regression, $1 < p \leq \infty$ is 0.5, 0.5, while for $L_1$ the regression values can be $\alpha,\, \alpha$ for any $0 \leq \alpha \leq 1$. 
Thus arbitrary binary splitting of the function values does not always produce the correct result when $p \neq 1$.

Instead one must split among the possible regression function values, not the original function values.
One important aspect is that for $L_p$ the weights need to be adjusted.
Before, when using splitting values $a,b$, with no function values in $(a,b)$, a function value $f(v)\leq a$ was temporarily changed to $a$, with the same weight because it was proportional to the increase in error if the regression value at $v$ was $b$ instead of $a$.
For $L_p$, $p > 1$ the change in error is not just $b-a$.
In~\cite{QPartition} it is shown that using weights based on the derivative of the error, $w(v)(a-f(v))^{p-1}$, gives the correct partitioning and can be used iteratively, just as before.

Here by approximation to within $\delta$ means that the regression values are no more than $\delta$ away from their value in an optimal isotonic regression.
For $1 < p < \infty$ there is only one optimal isotonic regression.
Rounding to a multiple of $\delta$ from the smallest function value, and splitting the range in half each time, gives a suitable approximation.

\begin{theorem}   \label{thm:Lpdeltaapprox}
Let $1 < p <\infty$.
Given a dag $G=(V,E)$, a weighted function $(f,w)$ on $V$ with integral weights in $[0,U]$, and $\delta >0$,
once a violator dag $\vio{G}=(\vio{V},\vio{E})$ has been constructed an isotonic function within $\delta$ of the $L_p$ isotonic regression can be found in $O(\mathcal{F}_2(\vio{n}, \vio{m}, nU) \log K)$ time,
where $K = \left(\max_{v \in V} f(v)  - \min_{v \in V} f(v)\right)/\delta$ and where the implied constants in the O-notation depend on $p$.
~$\Box$
\end{theorem}
Using the BLLSSSW flow algorithm gives $\tilde{O}(\vio{m})$ time if $K$ grows no more than polynomially in $n$ (a common assumption in this setting).
One can slightly improve the resulting regression by using the true $L_p$ mean and rounding to the nearest integer as the regression value for a level set.

Others have looked at approximation algorithms for all $L_p$ as well.
Kyng, Rao, and Sachdeva~\cite{Yale} give an algorithm which produces an isotonic regression within an additive error of $\delta$ from optimal, where they used total error, not vertex-wise.
It's moderately easy to adjust the $\delta$ in the above theorem to achieve their goal.
Their algorithm isn't based on an explicit violator dag, and takes $\tilde{\Theta}(m^{1.5} \log^2 n \log (nU/\delta))$ time to produce an appropriate approximation with probability at least $1-1/n$.

There is a more subtle way to do $L_p$ partitioning, but in a restricted setting.
For $L_1$ isotonic regression the values of each level set are a weighted median of the function values corresponding to the vertices in the set.
For $L_p$, finding splitting values that are guaranteed to separate the level sets is difficult since it isn't known how close these values can be to each other.
However, for integer function values and weights one can compute a $\delta$ such that no two level sets have values closer than $\delta$.
The details are in~\cite{QPartition}.
While the following theorem only analyzes binary values for arbitrary $p$, for $L_2$ the values can be arbitrary integers.
For $L_2$, one can use $\delta = 1/[\sum_{v \in V} w(v)]^2$, and no two subsets with different weighted means (their optimal regression value) can have weighted means closer than $4 \delta$, and hence calculating the mean of each level set gives the correct value.

The following result essentially appeared in~\cite{QPartition} but in terms of partitioning using the best $L_1$ algorithm known at that time.
Using a violator graph and the BLLSSSW flow algorithm results in a slightly faster algorithm.
Once the level sets have been determined the final step is to compute their $L_p$ mean, not just an approximation to within $\delta$,  and use that as the regression value.

\begin{corollary} \label{thm:Lpexact}
For $L_2$, if function values and weights are integers in
$[0,U]$, then for dag $G=(V,E)$ the exact isotonic regression can be found in the 
time given in Theorem~\ref{thm:Lpdeltaapprox}, where $\log K$ is replaced by $\log nU$.

For $L_p$, $1 < p < \infty$, the exact $L_p$ isotonic regression can be found in the same time if function values are in $\{0,1\}$ and integer weights are in $[0,U]$, where the implied constants depend on $p$.
~$\Box$
\end{corollary}

\section{Multidimensional and Related Orderings} \label{sec:multidim}

This section considers classes of orderings for which $T(\mathcal{V}(G)) = O(n^2)$, i.e., the violators can be found more quickly than for arbitrary dags and thus creating the violator dag might no longer be the bottleneck.
A very important partial ordering is given by points in $d$-dimensional space with coordinate-wise ordering.
Given points $x = (x_1,\ldots,x_d)$, $y=(y_1,\ldots,y_d)$, $x \neq y$, $y$ \textit{dominates} $x$ iff $x_i \leq y_i$ for $1 \leq i \leq d$.
Domination is also known as multi-dimensional ordering.
There is no requirement that the dimensions are the same, nor even numeric, merely that each is linearly ordered.
Data analytics involves rapidly increasing amounts of data where many datasets involve such orderings, often with a large number of vertices and occasionally with a significant number of dimensions. 
In Section~\ref{sec:Steiner} we show how to construct a violating graph which has more than $n$ vertices, but far fewer than $\binom{n}{2}$ edges, and which can be constructed quickly.
Theorem~\ref{thm:multidim} shows that this allows one to find an isotonic regression far faster than for arbitrary dags.

In Section~\ref{sec:pairwise} we give a class of graphs where $T(\mathcal{V}(G)) = \Theta(n^2)$, and in Section~\ref{sec:linearetal} we discuss linear and 2-dimensional orderings where $L_1$ and $L_2$ isotonic regressions can be constructed directly, without use of violators or flow algorithms.
These algorithms are well-known and for most no improvements are given here.

 The points may form a grid, which has nice properties such as the fact that the number of edges is linear in the number of points (throughout we ignore constants that depend on $d$, such as the number of edges emanating from a point in a grid, but will include factors such as $\log^d n$).
For 2-dimensional grids an $L_2$ isotonic regression algorithm, with no restrictions on values nor weights, appeared in~\cite{SpougeWanWilbur}, taking $\Theta(n^2)$ time.
An unrestricted $L_1$ algorithm, taking $\Theta(n \log n)$ time, appeared in~\cite{QPartition}.
Both algorithms used a scan-based dynamic programming approach, with the $L_1$ algorithm also incorporating partitioning.
By using trees to aid in the scan these can be extended to arbitrary points in 2-dimensions,
resulting in $\Theta(n \log^2 n)$ time~\cite{QPartition}.
However, the approach used in~\cite{SpougeWanWilbur} takes $\Theta(n^3)$ time.
 
 Violator-based algorithms do not improve these algorithms because while the original order is specified by a grid, a violator dag rarely would be.
 The $L_2$ algorithm is slower than that for $L_1$ because it can only guarantee that at each step it partitions into a set of 1 vertex and everything else in the other even though it scans through all vertices. However, in practice it is far faster than its worst case.

 \subsection{Steiner Vertices}\label{sec:Steiner}
 
Unfortunately the dynamic programming options in 2 dimensions do not extend to higher ones, so new approaches are needed.
In~\cite{QMultidim} a concise violator dag is given for domination ordering.  It incorporates Steiner vertices, which are vertices not in the original graph.
The coordinates in each dimension are converted to integers, where if a dimension has $x$ different values the coordinates are the integers in $[0, x-1]$.
They are represented as bit strings of $k$ bits, where $k= \lceil \lg x \rceil$.
Steiner  coordinates corresponding to strings of length $k$ are of the form
 ****, 0***, 1***, 00**, 01**, 10**,  \ldots 1110, 1111 (for $k=4$).
Given a vertex coordinate $q = q_1\cdots q_k$ and Steiner coordinate $t=t_1 \cdots t_k$,
$q \preceq t$ iff $q=t$ or there is a $j$, $1 \leq j < k$ such that $q_1\!=\!t_1$, $q_2\!=\!t_2 $,\ldots $q_j\!=\!t_j$, $q_{j+1}\!=\! 0$, $t_{j+1}\!=$*; and $t \preceq q$ iff $q=t$ or there is a $j$, $1 \leq j < k$ such that $q_1\!=\!t_1$, $q_2\!=\!t_2 $,\ldots $q_j\!=\!t_j$, $q_{j+1}\!=\!1$, $t_{j+1}\!=$*.
For vertex coordinates $q = q_1\cdots q_k$ and $r = r_1 \cdots r_k$, $q \neq r$, if $q \prec r$ then there is a $0 \leq j < k$ such that $q_i = r_i$ for $1\leq j$ and $q_{j+1}=0$, $r_{j+1}=1$.
Thus the Steiner coordinate $t = q_1 \cdots q_{j}$*$\cdots$* is such that $q \prec t \prec r$.

In the violator dag $\vio{G}=(\vio{V},\vio{E})$, a Steiner vertex $s$ is of the form $(s_1, \ldots s_d)$, where all the $s_i$ have Steiner coordinates and at least one has a *.
Let $S$ represent all Steiner vertices.
Then $\vio{V} = V \cup S$.
Given $p = (p_1,\ldots,p_d) \in V$ and $s=(s_1, \ldots, s_d) \in S$,
 $p \prec s$ in $\vio{G}$ iff $p_i \preceq s_i$ for all $1 \leq i \leq d$; and
 $s \prec p$ iff $s_i \preceq p_i$ for all $1 \leq i \leq d$.
 Add a directed edge $s \rightarrow v$ to $\vio{E}$ iff $v \prec s$, and $v \rightarrow s $ iff $s \prec v$.
There are no edges with both endpoints in $V$ or both in $S$.
For $p, q \in V$,  if $p \prec q$ as $d$-dimensional points (i.e., in the order we are constructing a representation of) then for each dimension $i$ let $s_i$ be the unique Steiner index such that $p_i \preceq s_i \preceq q_i$, and let $s=(s_1,\ldots,s_d)$.
If $p \neq q$ then there is at least one dimension $i$ where $p_i \neq q_i$, and hence $s_i$ has at least one *, i.e., $s$ is a Steiner point.
Thus $s$ is the unique Steiner point such that the edges $s \rightarrow v$ and $q \rightarrow s$ are in $\vio{E}$, and if $p \not\prec q$ there is no path in $\vio{G}$ from $q$ to $p$.
The property that for all $p, q \in V$, $p\prec q$ iff there is a path no longer than 2 from $q$ to $p$ is known as a \textit{2-transitive closure}, and adding the Steiner points makes it a \textit{Steiner 2-transitive closure}.
These have been studied for numerous reasons, see~\cite{2TransClosure}. 

Any vertex has at most $\prod_{i=1}^d k_i \leq \lceil \lg n\rceil^d$ incoming and outgoing edges, hence the total number of edges is $O(n \log^d n)$, and there are no more than this many vertices that have at least one edge.
In~\cite{QMultidim} it is shown that this dag, eliminating any Steiner points with only 0 or 1 incident edges and any edges that were connected to such Steiner points, can be created in $\Theta(n \log^d n)$ time, where the implied constants depend on $d$.
In~\cite{QMultidim} it is called a \textit{rendezvous graph}, and it is also shown that one dimension can be simplified, reducing the number of vertices, number of edges, and time to construct to $\Theta(n \log^{d-1} n)$.
This is called the \textit{reduced rendezvous graph}.
While some have created the equivalent of the rendezvous graph for use as a 2-transitive closure (see~\cite{2TransClosure} for relevant examples and references), the reduced rendezvous graph is not 2-transitive and hasn't been used in this community.

A simple modification can create a violator dag~\cite{QL0}:
add an extra dimension to the points and create the reduced rendezvous graph for them.
The value in this dimension is the function value.
Reversing the ordering of domination in this dimension, or making the value the negative of the function value and keeping the same definition, turns it into a violator dag.
Flow algorithms put weight 0 on the Steiner points.
Thus the time to construct the violator dag is $\Theta(n \log^d n)$ and $\hat{n}, \hat{m} = \Theta(n \log^d n)$.
Note that in general one cannot reduce the violator graph, removing vertices that aren't in any violators, any faster than constructing the violator graph.
Thus one does not know the true value of $\hat{n}$ until $\Theta(n \log^d n)$ time has occurred.
 
Combined with Theorem~\ref{thm:L0L1} this gives:

\setcounter{thmlistcounter}{0}
\begin{theorem} \label{thm:multidim}
Given a set of $n$ points in $d$-dimensional space, $d \geq 2$, with coordinate-wise ordering and integer weighted function $(f,w)$ on the points, with function values and weights in $[0,U]$:
\begin{list}
{\alph{thmlistcounter})}{\usecounter{thmlistcounter}\setlength{\leftmargin}{0.3in}}
\item an $L_1$ isotonic regression can be found in $O(\mathcal{F}_2(\vio{n}, \vio{n}, nU )\log n)$ time, where $\vio{n}=n \log^d n$ and the implied constants depend on $d$.
Using the CLLPPS flow algorithm gives $O(n^{1+o(n)}\log nU)$ time with high probability.
\item the same generic time bound applies for finding $L_0$ isotonic regression, with the final $\log n$ factor removed.
Using the CLLPPS flow algorithm has the same time, in O-notation, as above, also with the caveat that it is with high probability.
\end{list}
~$\Box$
\end{theorem}

For $d > 2$ and a weighted function the fastest previous $L_1$ algorithms took $\Theta(n^2 \log^{d} n)$ time, and $\Theta(n^{1.5} \log^{d+1}n)$ for unweighted functions~\cite{QPartition}.

Combining Theorems~\ref{thm:multidim} and~\ref{thm:Lpdeltaapprox} gives

\begin{corollary}
Given a set of $n$ points in $d$-dimensional space, $d \geq 2$, with coordinate-wise ordering, given a weighted function $(f,w)$ on the points, given $p \in (1,\infty)$, and $\delta > 0$, an isotonic regression within $\delta$ of the $L_p$ isotonic regression can be found in $O(\mathcal{F}(\vio{n}, \vio{n}, nU) \log K)$ time, where $\vio{n}=n \log^d n$, $K = \left(\max_{v \in V} f(v)  - \min_{v \in V} f(v)\right)/\delta$ and the implied constants in the O-notation depend on $p$ and $d$.

Using the CLLPPS flow algorithm gives $O(n^{1+o(n)}\log K \big)$ time with high probability, where the implied constants are dependent on $p$ and $d$.~$\Box$
\end{corollary}

\subsection{Orderings Given By Pairwise Comparisons} \label{sec:pairwise}

Some problems may not initially look like they can be framed as multi-dimensional vertices.
For example, suppose we collect estimates of the number of people within a range of income and age, for various ranges.
This is asking for estimates of the number of people in an axis-parallel (iso-oriented) rectangular region, and the correct values are isotonic in that if one range contains another the former has at least as many people as latter (i.e., there are no people with negative weight).
A rectangle $R$ with lower left coordinates $(x_1^r,y_1^r)$ and upper right coordinates $(x_2^r,y_2^r)$ is contained in a rectangle $S$ with coordinates $(x_1^s,y_1^s)$ and $(x_2^s,y_2^s)$ iff $(x_1^s,y_1^s) \preceq (x_1^r,y_1^r)$ and
$(x_2^r,y_2^r) \preceq (x_2^s,y_2^s)$. 
By reversing the ordering on two dimensions, this is just a 4-dimensional ordering, and the same technique extends to axis-aligned boxes in arbitrary dimensions.
Sets of axis-parallel boxes have been studied in other settings, such as generating all intersecting pairs, but there don't seem to be previous algorithms which represent all violating pairs so concisely in a dag.

A slight variant of this does not seem to have such a simple multidimensional representation, but does have the special property that the vertices are objects of some form (such as the $d$-dimensional coordinates above) and the ordering is given explicitly by direct comparisons of objects, rather than via a dag.
For example, one may have a collection of rectangles in arbitrary orientation, not necessarily all axis-parallel, and want estimates of their area, where the ordering is still containment.
While there isn't a Steiner-based representation to reduce the size of a violator dag (to the author's knowledge), direct pairwise comparisons allows one to determine the transitive closure and create the violator dag in $\Theta(n^2)$ time, with $\hat{n}=n$ and $\hat{m} \leq \binom{n}{2}$.
As for the multidimensional points, one does not the true value of $\hat{n}$ nor $\hat{m}$ until the dag has been constructed.

Additional examples with the property of being based on vertices which are pairwise compared are
\begin{itemize}
\item strings over some alphabet where $v \prec w$ iff $v$ is a substring of $w$
\item strings where $v \prec w$ iff $v$ is a subsequence of $w$
\item convex polyhedra in $d$-dimension space where the ordering is by containment
\item subsets of a given set, with ordering by containment
\item positive integers, where $v \prec w$ iff $v$ evenly divides $w$
\end{itemize}
For some of these examples the time to do comparisons does not have a fixed bound and would need to be incorporated in the time analysis.

\subsection{Planar Dags}

Planar dags of $n \geq 3$ vertices have at most $3n-6$ edges, and thus are similar to the pointwise comparison graphs in that the transitive closure can be computed in $\Theta(n^2)$ time, and hence the violator graph can be computed in this time.
One slight difference is that here the ordering is given by an explicit dag, i.e., is given in the same format as an arbitrary ordering given via dag. The overall structure of the approach given in Section~\ref{sec:background} states that the dag  being evaluated has already had the non-violating vertices and their incident edges removed and hence $\hat{n}= n$, $\hat(m) \leq \binom{n}{2}$.

Note that one does not need an explicit embedding of the vertices and edges into the plane, merely the knowledge that it can be done.

\subsection{Linear and 2-Dimensional Orderings}  \label{sec:linearetal}

Nothing in this paper improves on previously known results for points in 1- or 2-dimensional orderings, other than the 2-dimensional results for $L_0$, a combination that has apparently never been analyzed.
For linear orders with arbitrary real values and weights the PAV (pool adjacent violators) approach gives a simple, well-known,
 $\Theta(n)$ algorithm for $L_2$ and more complicated $\Theta(n \log n)$ algorithms for $L_1$~\cite{AhujaOrlin}.
 Partitioning gives a simpler $\Theta(n \log n)$ algorithm for $L_1$.
 For $L_0$ the simple longest increasing subsequence algorithm (LIS) can be used for unweighted functions, and when modified to find the sequence with the largest total weight (which isn't necessarily longest) finds an optimal regression in $\Theta(n \log n)$ time.
Nothing here improves upon these times.

For $d=2$ dynamic programming approaches are known for $L_1$ and $L_2$, where for a 2-d grid a $\{0,1\}$-valued $L_1$ regression can be found in $\Theta(n)$ time, and hence for arbitrary values and weights $L_1$ isotonic regression can be solved in $\Theta(n \log n)$ time via partitioning~\cite{QPartition}.
For $L_2$ the time of the dynamic programming algorithm in~\cite{SpougeWanWilbur} is $\Theta(n^2)$.
Neither algorithm involves violator graphs and each gives an exact result.
Since on a 2-d grid the $\{0,1\}$-valued $L_1$ regression can be determined in $\Theta(n)$ time, 
it can be used to replace the flow approach used in Corollary~\ref{thm:Lpexact}, giving the 
result in Theorem 5.6 of~\cite{QPartition}, namely that if integral function values and weights are in
$[0,U]$ then for a 2-dimensional grid of $n$ vertices the exact $L_2$ regression can be found in $\Theta(n \log U)$ time,
and for 2-d points in general position can be found in $\Theta(n \log n \log U)$ time.
Unfortunately, a cursory survey of implementations mentioned on the web shows that for 2-d grid orders the slower Spouge, Wan, and Wilbur~~\cite{SpougeWanWilbur} algorithm is the one that is used overwhelmingly.

For $d \geq 3$ no dynamic programming approaches are known, and thus having the points arranged in a grid does not give faster time than having the points just be in arbitrary positions.

\section{Final Remarks} \label{sec:final}

\begin{table}[p]
\begin{center}

\large{$L_0$} Exact, $\ell$ different function values
\vspace{0.08in}

\begin{tabular}{|l|cc|cc|} 
 \multicolumn{1}{c}{}  & \multicolumn{2}{c}{previous best} & \multicolumn{2}{c}{best here, integer values and weights} \\ \hline
 graph      & time & reference & time &  reference \\ \hline
 linear       & $\Theta(n \log \ell)$   & LIS  & no improvement  & \\ 
                     & \multicolumn{2}{c|}{real values, weights}           &    \multicolumn{2}{r|}{}                    \\ \hline
$d \geq 2$   \up&   $\Theta\left(n^3\right)$  & A & $\tilde{O}\left( (n \log^{d} n)^{1+o(n)}\log nU\right)$ & Thm~\ref{thm:multidim} \\ \hline
pairwise compare \up&   $\Theta\left(n^3\right)$  & A  & $\tilde{O}(n^2+\hat{m} \log nU)$   & Thm~\ref{thm:L0L1} \\ \hline
 arbitrary          \up                        &  $\Theta\left(n^3\right)$ &  \cite{Relab_Feelders06,Relab_PijlsPot14,Relab_Radeeta09}~ &   $\tilde{O}\left(\min\{nm, n^\omega\}~+ ~~~~~~~~\right.$ & Thm \ref{thm:L0L1}   \\
 & &  & \multicolumn{2}{c|}{$\left.\vio{m}\log U \right)~~~~~~~~$} \\ \hline

\end{tabular}
\vspace{0.30in}

\large{$L_1$} Exact, $\ell$ different function values
\vspace{0.08in}

\begin{tabular}{|l|cc|cc|}
 \multicolumn{1}{c}{}  & \multicolumn{2}{c}{previous best, real values \& weights} & \multicolumn{2}{c}{best here, integer values \& weights} \\ \hline
  graph      & time & \hspace{-0.5in} reference & time &  \hspace{-0.5in}reference \\ \hline
  linear  \up & $\Theta(n \log \ell)$  & \hspace{-0.3in} partition or PAV & no improvement & \\ \hline
   $d = 2$ grid  \up & $\Theta(n \log \ell)$      & \cite{QPartition}  & no improvement   & \\ \hline
      $d = 2$ arb   \up & $\Theta(n \log n \log \ell)$      & \cite{QPartition}  & no improvement   & \\ \hline
$d \geq 3~~ \cite{QMultidim}$     \up & $\Theta\left(n^2 + \hat{n}^2 \log^{2d+1} \hat{n} \right)$  &  A   &   $\tilde{O}\left( n \log^d n~+ \hspace*{0.8in} \right.$ & Thm \ref{thm:multidim} \\
  & \multicolumn{2}{c|}{ } & \multicolumn{2}{c|}{$\left.(\hat{n} \log^{d} \hat{n})^{1+o(\hat{n})}\log U \log \hat{n}\right)$~~~} \\ \hline
pairwise   \up &  $\Theta\left(n^2 + \hat{n}\hat{m} + \hat{n}^2 \log n \right)$  &  A  & $\tilde{O}(n^2+\hat{m}\log U \log n)$ & Thm \ref{thm:L0L1}   \\ \hline
arbitrary          \up                        &  $\Theta\left(nm + n^2 \log n\right)$ &  \cite{AHKW} &   $\tilde{O}\left(\min\{nm, n^\omega\}~+ \right.$ & Thm \ref{thm:L0L1}   \\
 & &  & \multicolumn{2}{c|}{$\left.\vio{m} \log U \log n\right)$~~~~~~~~~} \\ \hline

\end{tabular}
\vspace{0.30in}

\large{$L_2$}  Exact
\vspace{0.08in}  

\begin{tabular}{|l|cc|cc|} 
   \multicolumn{1}{c}{}  & \multicolumn{2}{c}{previous best, real values \& weights} & \multicolumn{2}{c}{best here, integer values \& weights} \\ \hline
    graph      & time & reference & time &  reference \\ \hline
  linear  \up & $\Theta(n)$  & PAV  & no improvement & \\ \hline
  $d=2$ grid  \up &  $\Theta(n^2)$  &  \cite{SpougeWanWilbur}  &  $\Theta(n \log U)$  &  \cite{QPartition}  \\ \hline
  $d=2$ arb   \up&  $\Theta(n^2 \log n)$   &  \cite{QPartition}  &  $\Theta(n \log n \log U)$   & \cite{QPartition}  \\ \hline
     $d \geq 3$~~ \cite{QMultidim}    \up &  $\tilde{\Theta}\left(n^2 + \hat{n}^2 \log^{2d+1} \hat{n}\right) $  &  A &  $\tilde{O}\left((n \log^d n)^{1+o(n)}\log nU\right)$  &  Thm \ref{thm:multidim}  \\ \hline
  pairwise   \up&  $\Theta\left(n^2 + \hat{n}\hat{m} \log \frac{\hat{n}^2}{\hat{m}} \right)$  &  A &  $ \tilde{O}(n^2+ \hat{m}\log nU)$   &  Thm 2 \\ \hline
   arbitrary          \up                        &  $\Theta\left(nm \log \frac{n^2}{m}\right)$ &  \cite{HochQuey} &   $\tilde{O}\left(\min\{nm, n^\omega\}~+ \right.$ & Thm \ref{thm:L0L1}   \\
 & &  & \multicolumn{2}{c|}{~~$\left.\hat{m} \log nU\right)$~~~~~~~~~} \\ \hline
\end{tabular}
\smallskip
\end{center}
\caption{Previous fastest algorithms vs.~approaches here using currently fastest algorithms} 
\label{tab:oldvsnew}
\bigskip

``A'' means that the algorithm for arbitrary dags is used, coupled with $T(\mathcal{V}(G))$.
\medskip

Previous algorithms are worst-case times, while here they are expected time with high probability.
All timings assume standard operations take unit time, even on numbers as large as $U$.
\smallskip

To accommodate future improvements, or to evaluate time when something suboptimal is used, go back to the theorems and use the appropriate values for $\mathcal{V}$ and $\mathcal{F}$.
\end{table}

Isotonic regression is a useful non-parametric shape-constrained approach to problems where assumptions such as linearity, logistic, or other parametric requirements are dubious.
It has growing popularity in data analysis and machine learning \cite{MonotonicLattices,LearnRelab_Cano18,MonoNeuralNet,DuivFeeldersNNclass,Sparse,Isotron,Certified,Relab_Radeeta09,TaoWang21}, but to scale to the future size of some of these problems requires faster algorithms.
This paper describes such algorithms without restricting to any specific application.
Eventually parallel algorithms will be needed, but they have had little attention so far.

 Recent advances in maximum flow algorithms~\cite{Flow_BLLSSSW,Flow_CKLPPS}, coupled with the standard approach for $L_0$ isotonic regression and $L_1$ partitioning, gives faster serial algorithms for exact solutions for $L_0$, $L_1$, and $L_2$ when the weights are positive integers in $[0,U]$ (and for $L_2$  the function values are similarly constrained).
 For $L_p$, $1 < p < \infty$, we give algorithms for approximate~\ref{thm:Lpdeltaapprox} and exact~\ref{thm:Lpexact} isotonic regressions, where the latter has severe restrictions on the values.
Table~\ref{tab:oldvsnew} summarizes the times of the algorithms developed here vs.\ the fastest ones published previously among those that produce exact answers (to within machine round-off error) for real-valued input.
It does not include the approximation algorithms in Kyng, Rao, and Sachdeva~\cite{Yale} which are similar to ours but slightly slower.
For $d$-dimensional vertices theirs takes $\tilde{\Theta}(n^{1.5} \log^{1.5(d+1)})$ time vs.\ the $O(n^{1 + o(n)})$ here, and for an arbitrary dag takes $\Omega(n^3)$ vs.\ the $\Theta(\min\{nm,n^\omega\})$ here.

One drawback of using a violator graph is that in the worst case $\vio{m} =\Theta(n^2)$ no matter how small $m$ is.
Fortunately, for the very important class of $d$-dimensional component-wise ordering (domination) there is a violator dag where $\vio{m} = \Theta(n \log^d n)$ (Sec.~\ref{sec:multidim}).
The time and space to construct it are only a factor of $\log n$ more than constructing the domination graph, and it is somewhat sparse no matter were the points are located.
Similarly, if the graph is determined by pairwise comparison of vertex labels then the time and space is $\Theta(n^2)$ for both the underlying order and the violator graph.

When a violator dag is given the times are determined by the performance of the flow
 algorithm, so efficient implementations of these complex algorithms are needed.
 However, our time analyses are stated in general terms, so whatever violator graph and flow algorithms are used the time is given in terms of their performance.
 This accommodates advances in flow algorithms, and already shows improvements in isotonic regression algorithms using  recent flow algorithms.
It also accommodates analyzing the use of algorithms that are far from optimal.

Assuming $U$ grows at most polynomially in $n$, for arbitrary orderings determined by pairwise comparison of points' function values (integers in $[0,U]$) the previous worst case for $L_0$, $L_1$, and $L_2$ was $\Theta(n^3)$, while here it is $\tilde{\Theta}(n^2)$, which is probably within log factors of optimal even for sparse graphs since every possible edge needs to be generated.
For an arbitrary dag the previous best algorithms for $L_0$, $L_1$, and $L_2$ isotonic regression took $\Theta(n^3)$ time, while here they take $\Theta(n^\omega)$.
Thus this approach, utilizing the currently fastest flow algorithms, has apparently achieved the best possible for dense graphs since it isn't known how to find the transitive closure faster, and $\Theta(n^\omega)$ will be optimal as long as $\omega > 2$.
Usually this analysis is discussed in terms of the smallest known value of $\omega$, but here $\omega$ can be used to indicate the time of the algorithm actually used (such as $\omega \approx 2.81$ for Strassen's), not necessarily a value that requires a galactic size input before it is faster than far simpler algorithms.

\end{document}